\documentclass{aipproc}

\usepackage{graphicx} 

\newcommand{\Gray}{$\gamma$-ray\ }

\newcommand{\Cerenkov}{\v Cerenkov\ }
\newcommand{\emphasise}[1]{\textbf{#1}}

\layoutstyle{6x9}

\begin{document}

\title{VHE observations of unidentified EGRET sources}

\author{S.J. Fegan$^*$}{
address={Fred Lawrence Whipple Observatory, PO Box 97, Amado, AZ 85645, USA},
address={Physics Department, University of Arizona, Tucson, AZ 85721, USA},
email={sfegan@egret.sao.arizona.edu},
}

\author{the VERITAS collaboration}{
address={Fred Lawrence Whipple Observatory, PO Box 97, Amado, AZ 85645, USA},
}

\begin{abstract}
Observations of unidentified EGRET sources were made with the Whipple
10m imaging atmospheric \Cerenkov telescope between Fall 1999 and
Spring 2001. During this period, a high resolution 490 pixel camera
with $4^\circ$ field of view was present on the
telescope. Characterization of the off-axis response of this
instrument was done using observations of the Crab Nebula. No
significant emission was detected from the eight unidentified EGRET
sources observed and upper limits are presented as a function of
position.
\end{abstract}

\maketitle

\section{Introduction}

Very High Energy (VHE) \Gray astronomy is the term used to describe
observations in the energy range from 300GeV to 100TeV. 
Ground-based instruments
operating in this energy domain typically have large collecting areas,
good angular resolution and relatively large fields of view. 
The
atmospheric \Cerenkov imaging technique is described in detail
elsewhere \citep{Ong98}. 


Twenty five years of gamma ray observations in the MeV to GeV range,
have produced
nearly 300 cataloged sources. During its lifetime, the EGRET
experiment aboard CGRO made the most significant contribution to the list
of detected sources, although its relative insensitivity to the arrival
direction of 100 MeV photons means that the location of many sources are
only known to within $\sim0.5^\circ$.
The majority of sources are, as yet, not firmly associated
with objects at other wavelengths. In many cases the EGRET error
circle is populated by a number of prospective X-ray, optical and
radio sources which are all candidate associations.

\section{Observations}

Observations of eight unidentified EGRET sources, listed in Table
\ref{OBS::SOURCES}, were made with the Whipple 10m imaging atmospheric
\Cerenkov telescope in Arizona USA.  The instrument and its
characteristics are described in
\citet{Finley01}.


For off-axis and extended sources the telescope is operated in ON-OFF
mode. Each 28 minute scan of the source region is followed by a 28
minute control run offset from the source by 30 minutes in right
ascension and in time. Taking the control data in this manner
compensates for differences in brightness that are a function of
elevation and azimuth.


\begin{table}[ht]
\caption{\label{OBS::SOURCES} Unidentified EGRET sources selected for observation.}
\begin{tabular}{ccccccccc}

& \multicolumn{4}{c}{\emphasise{Position}} &
\emphasise{Observation} & \emphasise{Exposure} \\

\emphasise{Source} & 
\emphasise{RA} & \emphasise{Dec} & \emphasise{l} & \emphasise{b} & 
\emphasise{dates} & \emphasise{(min)} \\ \hline
3EG J0423+1707 & 04:23:00 & 17:06:60 & 178.48 & -22.14 & 2000/12 - 2001/02 & 248 \\
GeV J0433+2907 & 04:33:38 & 29:05:56 & 170.50 & -12.58 & 1999/11 - 2000/01 & 500 \\
3EG J0450+1105 & 04:50:00 & 11:05:00 & 187.86 & -20.62 & 2000/11 - 2001/01 & 274 \\
3EG J0634+0521 & 06:33:12 & 05:53:07 & 206.18 &  -1.41 & 2000/11 - 2001/03 & 275 \\
3EG J1323+2200 & 13:23:03 & 21:59:41 & 359.33 &  81.15 &  2001/01 - 2001/02 &  83 \\
GeV J1907+0556\tablenote{\citet{Roberts01} note that this source is over $1^\circ$
away from 3EG J1903+0550, with which it is associated in \citet{Hartman99}.
They conclude that this association is likely to be incorrect.} & 19:07:41 & 05:57:14 &  40.08 &  -0.88 & 2000/05 - 2000/06 & 277 \\
GeV J2020+3658\tablenote{This source is incorrectly associated with 3EG J2016+3657 in
the third EGRET catalog. \citet{Roberts01} note that 3EG J2021+3716 is
consistent with the GeV source.
} & 20:20:43 & 36:58:38 &  75.29 &   0.24 & 1999/10 - 1999/11 & 139 \\
3EG J2227+6122 & 22:27:14 & 61:22:15 & 106.53 &   3.18 & 2000/09 - 2000/10 & 341 \\ \hline
\end{tabular}
\end{table}


\section{Analysis}

Before any analysis is performed, all data are subject to a number of
standard operations. First, the data is flat-fielded, a process which
compensates for any non-uniformities in the camera. Second,
artificially generated noise is added to each image to remove any
biases that exist between the on-source and control observations, a
process referred to as \textit{software padding}. These biases result
from the control data being taken while pointing to a different part
of the sky which has different background light
characteristics. Finally, each image is cleaned by ignoring all
channels which do not have sufficient signal in them. The details of
these procedures are described elsewhere \citep{Reynolds93,Cawley90}.

For extended sources or sources where the source location is not well
determined, it is essential to reconstruct the arrival direction of the
primary. 
The arrival direction must be inferred from the ``shape'' of the
observed image. There are a number of methods available, the approach
taken here, described in detail in \citet{Lessard01}, is to assume
that the arrival direction of the primary lies along the major axis of
the shower image and is displaced from the center of the shower image
by a distance given by,
\begin{displaymath}
disp=\xi\left(1-\frac{width}{length}\right)
\end{displaymath}
where $width$ and $length$ describe the shape of the recorded image
and $\xi$ is a scaling parameter.

This method yields two possible arrival directions, each of which is
on the major axis of the shower image, seperated from the centroid by
the calculated parameter, $disp$.  When creating a 2D map the origin of each
event is assigned to both possible directions in the hope that one
will have an excess as more event origins are superimposed.

A sky map is then produced by building up a 2-dimensional histogram of
the reconstructed arrival direction with respect to the center of the
camera. Errors in reconstructing both the image axis and $disp$ are
accounted for by convolving the final 2D map with a Gaussian function
\(g(\vec{r};r_0)= \exp(-r^2/2r_0^2)\), where $r_0$ is a scaling parameter
chosen to maximise the significance of an excess.

Calculation of excess signal, significance and upper-limit maps
($S(\vec{r})$, $\sigma(\vec{r})$ and $UL(\vec{r})$ respectively) is
then done by convolving the ON and OFF counts with the smoothing
function $g(\vec{r})$ in the appropriate manner,
\(
S(\vec{r})=\sum_{\vec{r'}}{[ON(\vec{r'})-OFF(\vec{r'})]g(\vec{r'}-\vec{r})}
\) and 
\(
\Delta S(\vec{r})^2=\sum_{\vec{r'}}{[ON(\vec{r'})+OFF(\vec{r'})]g^2(\vec{r'}-\vec{r})}
\). 
Then $\sigma(\vec{r})=S(\vec{r})/\Delta S(\vec{r})$ and $UL(\vec{r})$
is calculated from $S(\vec{r})$ and $\Delta S(\vec{r})$ by the method
of \citet{Helene83}.

\section{Calibration}

Calibration of the two dimensional analysis method was done using sets
of observations of the Crab Nebula. 
Taking observations with the source location offset from the center
of the field of view by various degrees and calculating the relative
\Gray rate allows a model of the detector response for off-axis and
extended sources to be made.


\begin{figure}[t]
\begin{minipage}{0.33\textwidth}
\centerline{On axis}
\resizebox{\textwidth}{!}{\includegraphics{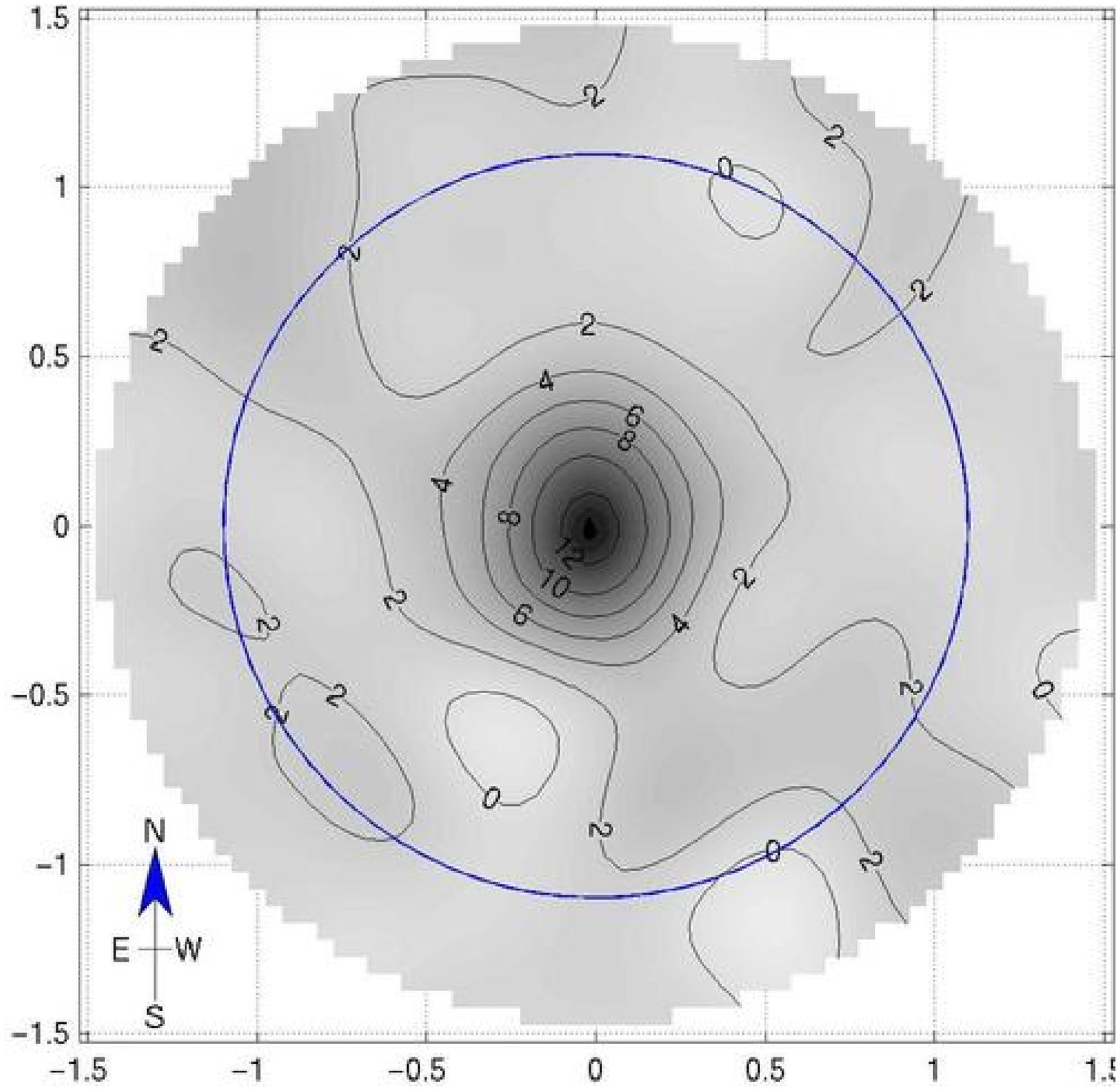}}
\end{minipage}
\begin{minipage}{0.33\textwidth}
\centerline{$0.3^\circ$ off axis}
\resizebox{\textwidth}{!}{\includegraphics{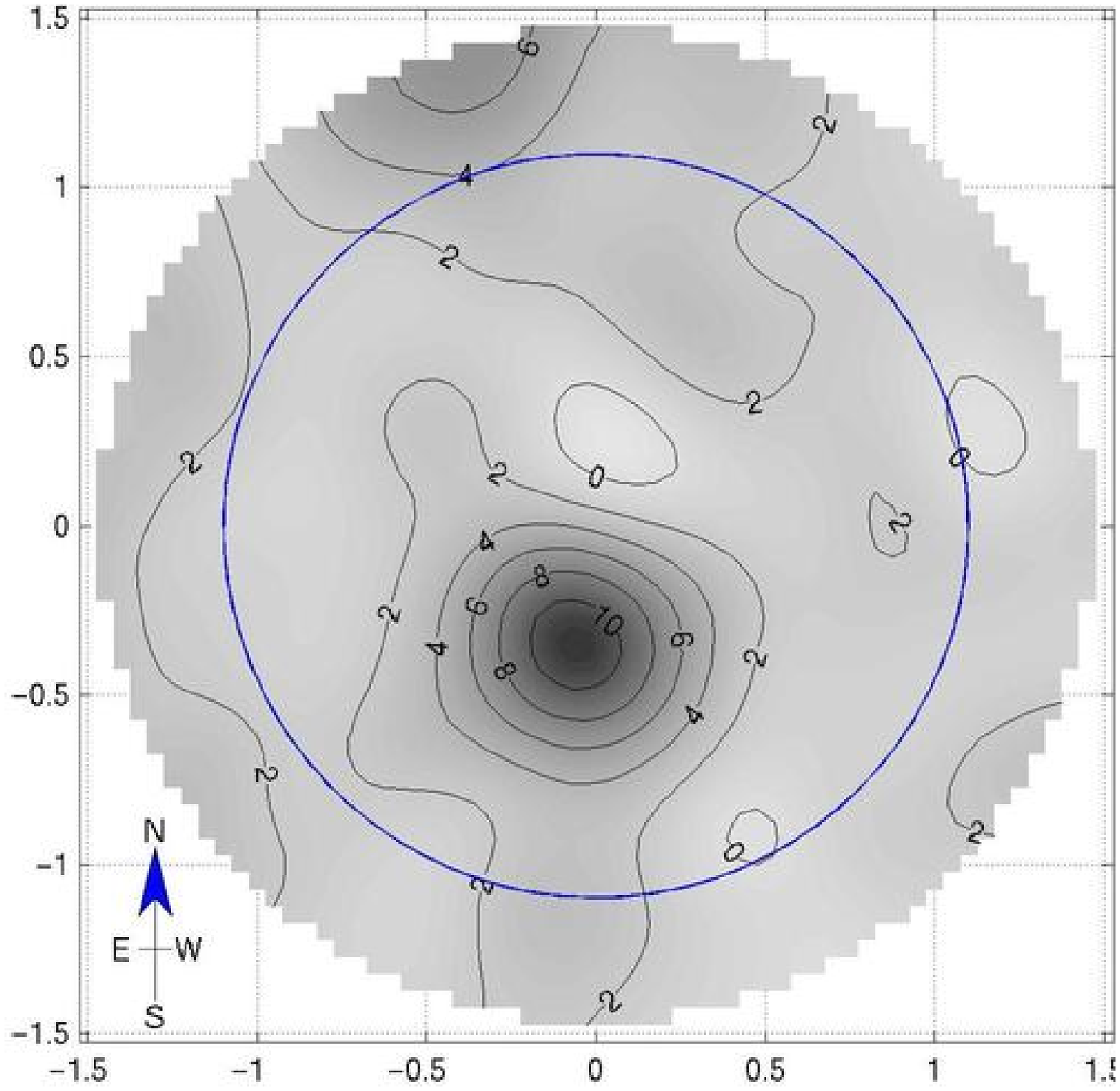}}
\end{minipage}
\begin{minipage}{0.33\textwidth}
\centerline{$1.3^\circ$ off axis}
\resizebox{\textwidth}{!}{\includegraphics{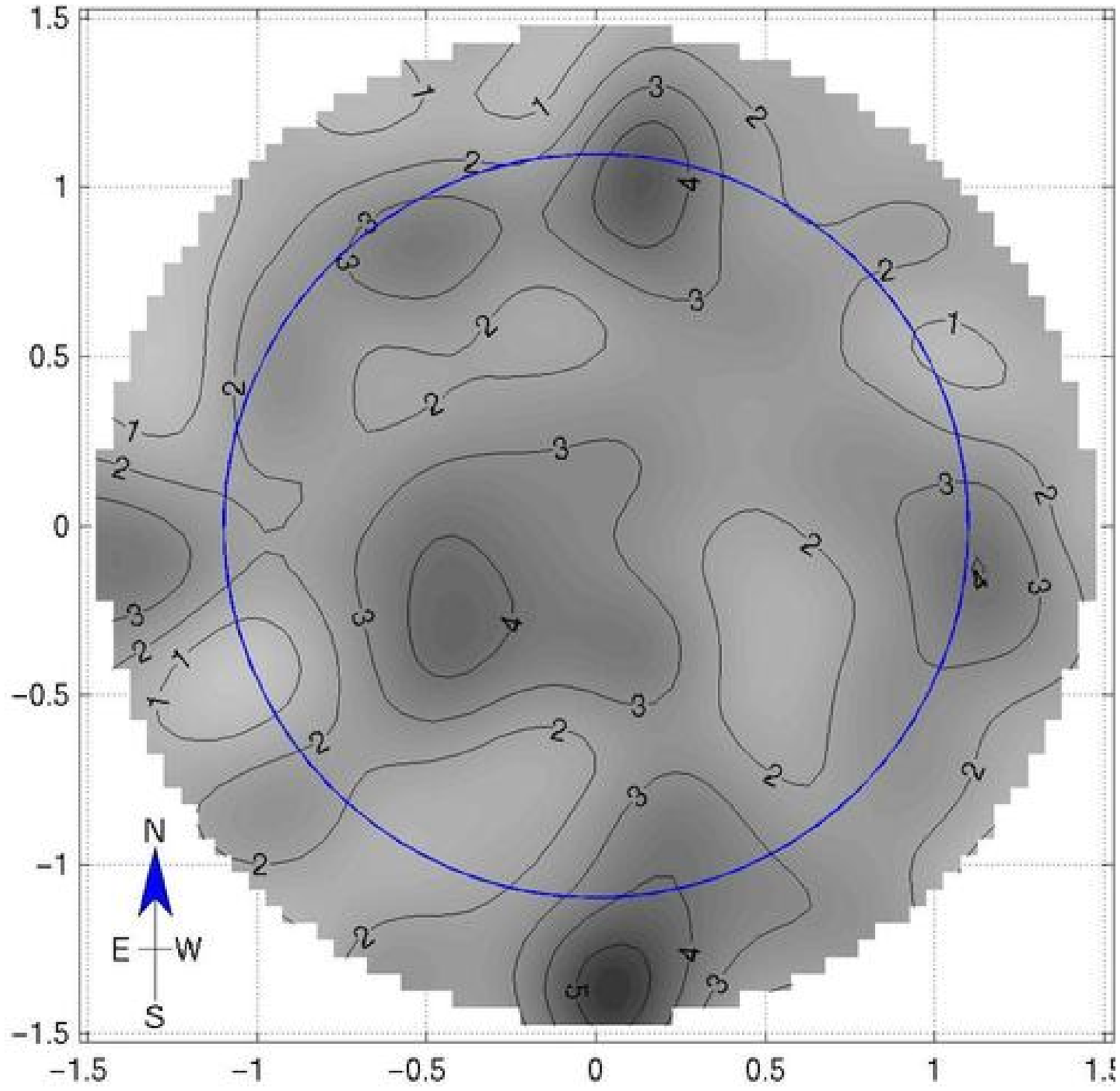}}
\end{minipage}
\caption{\label{CALIBRATION::SIGNIFICANCES} Observations of the Crab Nebula, 
offset by varying amounts from the center of the field of view. The
contours show detection significance. Positions are in degrees from
the center of the field of view with RA and Dec increasing toward the
left and top respectively. A circle of radius $1.1^\circ$ denotes the
geometrical extent of the camera used. The observations at an offset of
$1.3^\circ$ place the Crab outside of this.}
\end{figure}

Figure \ref{CALIBRATION::SIGNIFICANCES} shows significance maps for
the Crab Nebula offset by three different amounts. In each of them the
Crab is clearly visible. At an offset of $0.3^\circ$ the \Gray collection
efficiency is $84$\% of what it is on axis.  At an offset of
$1.3^\circ$, with the source outside of the geometrical extent of the
camera, the efficiency is $30$\%. The significance map for this data
shows appreciable background contamination over the field due to the
simple reconstruction approach of assigning the arrival direction of
each photon to two points on the shower axis. More sophisticated
approaches can reduce such false sources \citep{Lessard01}.

\begin{figure}[h]
\centerline{\resizebox{0.75\columnwidth}{!}{\includegraphics{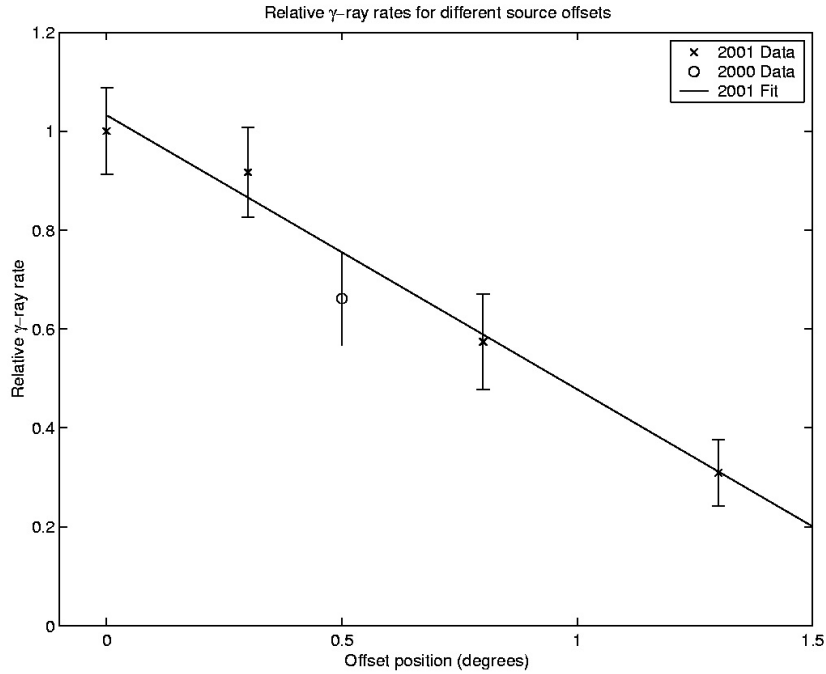}}}
\caption{\label{CALIBRATION::RATE} Relative Crab rate as a function of source
offset. The off-axis response can be fit by a straight line.}
\end{figure}

Figure \ref{CALIBRATION::RATE} shows the relative collecting
efficiency for offset sources. This curve is used to normalize
detected emission rates or upper limits to the Crab flux. 


\section{Results}
\begin{figure}[h]
\begin{minipage}{\textwidth}
\newlength{\plotwidth}
\setlength{\plotwidth}{0.32\textwidth}
\begin{minipage}{\plotwidth}
\centerline{3EG J0423+1707}
\resizebox{\textwidth}{!}{\includegraphics{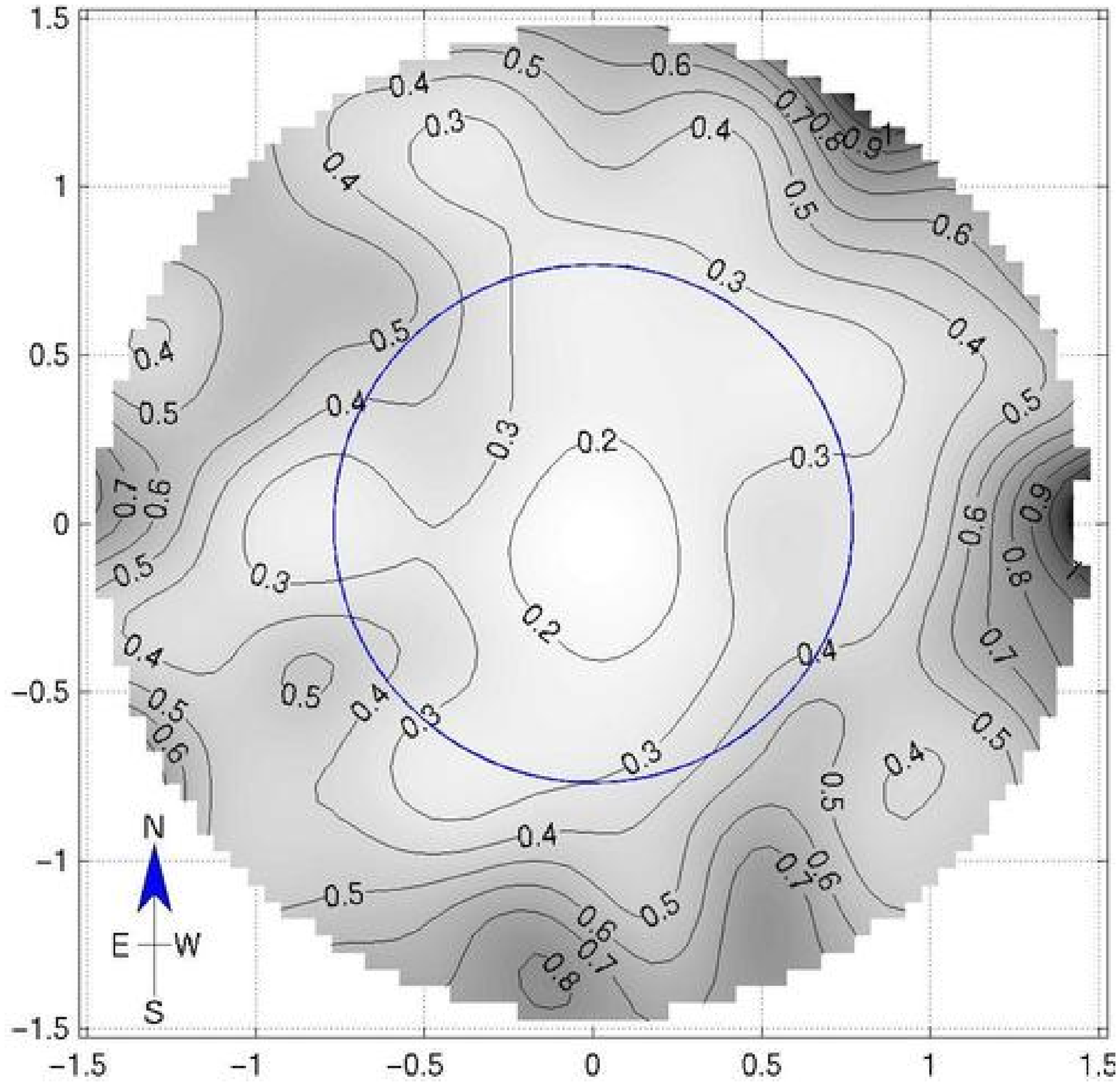}}
\end{minipage}
\begin{minipage}{\plotwidth}
\centerline{GeV J0433+2907}
\resizebox{\textwidth}{!}{\includegraphics{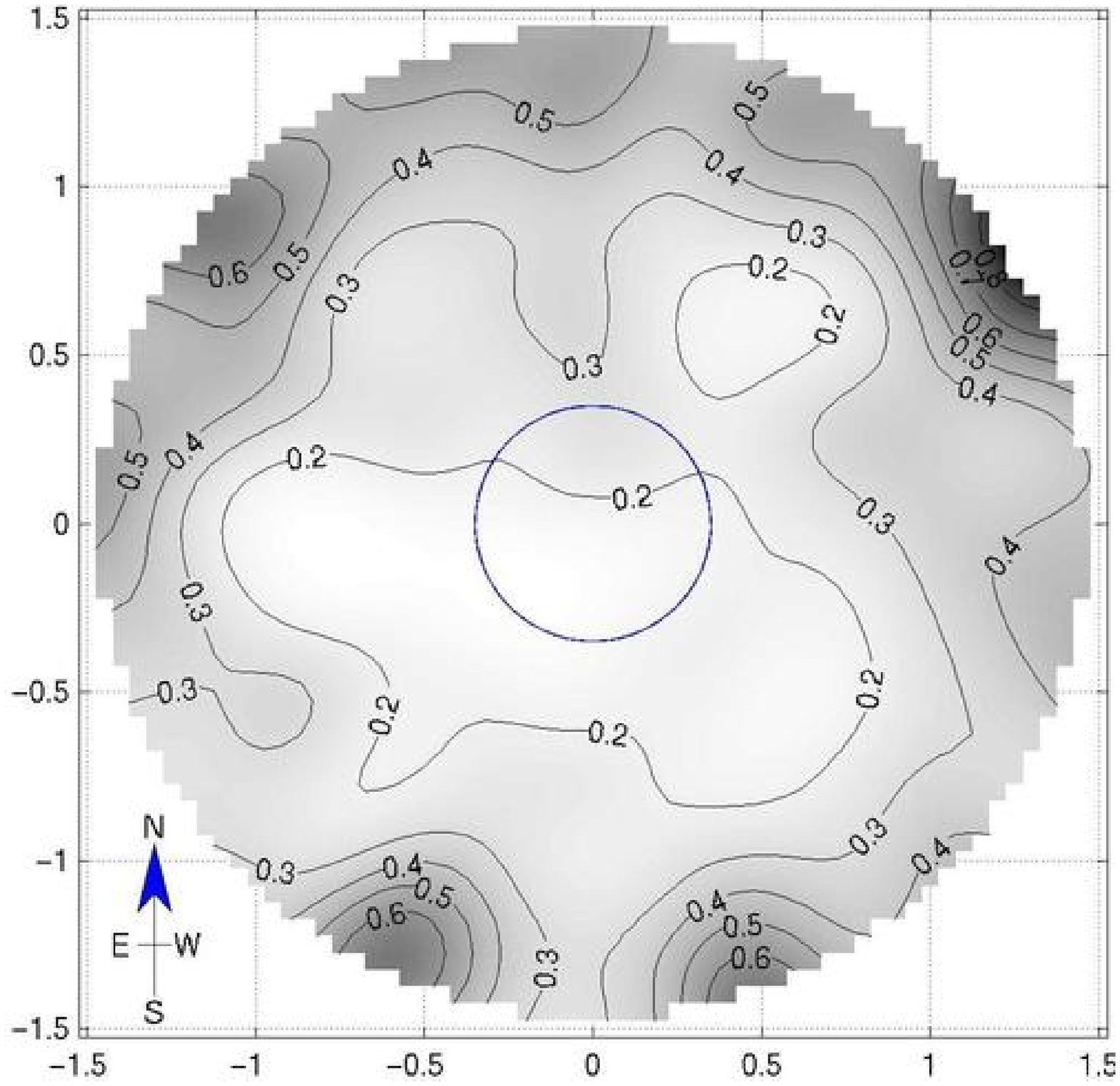}}
\end{minipage}
\begin{minipage}{\plotwidth}
\centerline{3EG J0450+1105}
\resizebox{\textwidth}{!}{\includegraphics{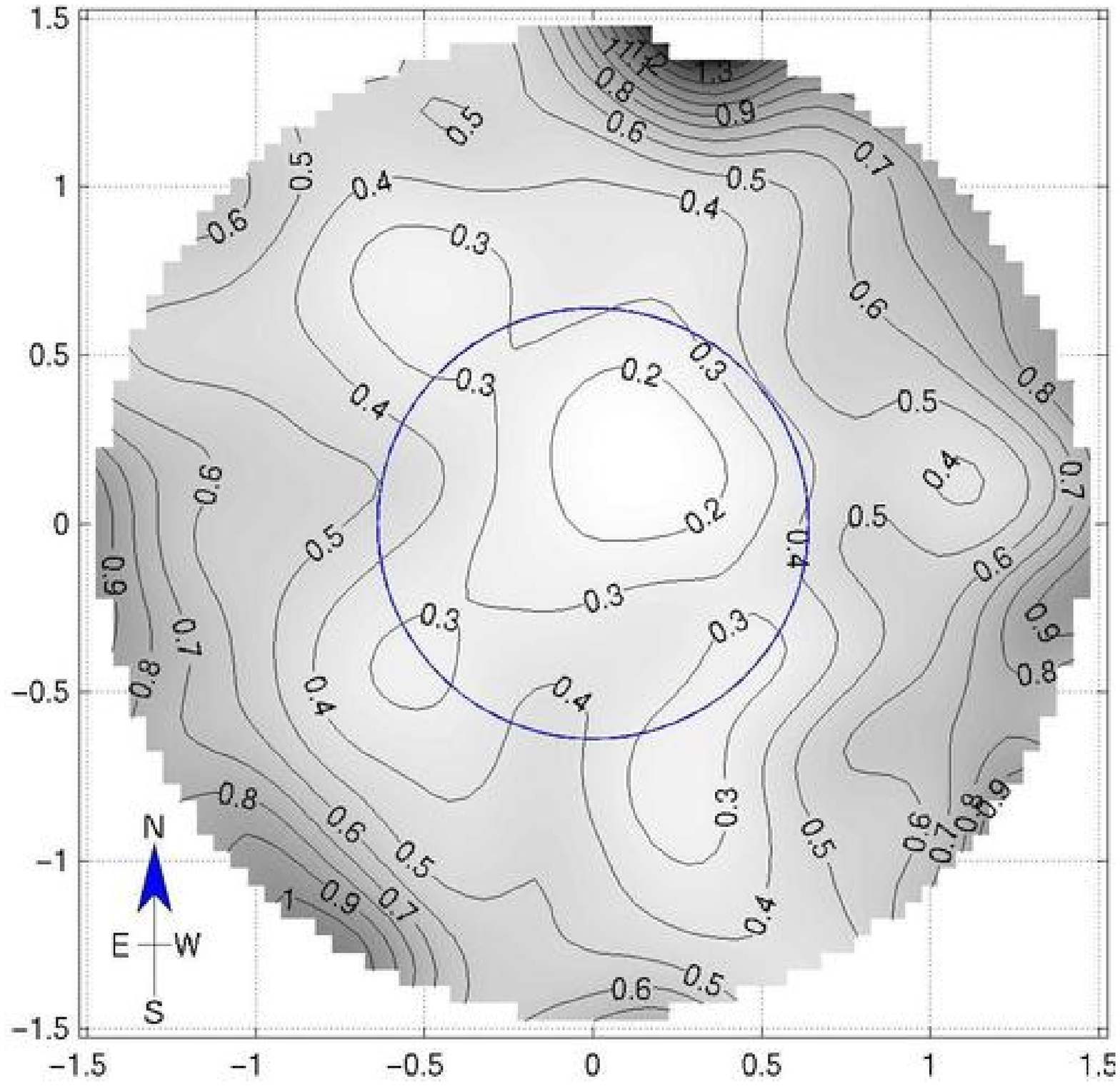}}
\end{minipage}

\vspace*{2ex}
\begin{minipage}{\plotwidth}
\centerline{3EG J0634+0521}
\resizebox{\textwidth}{!}{\includegraphics{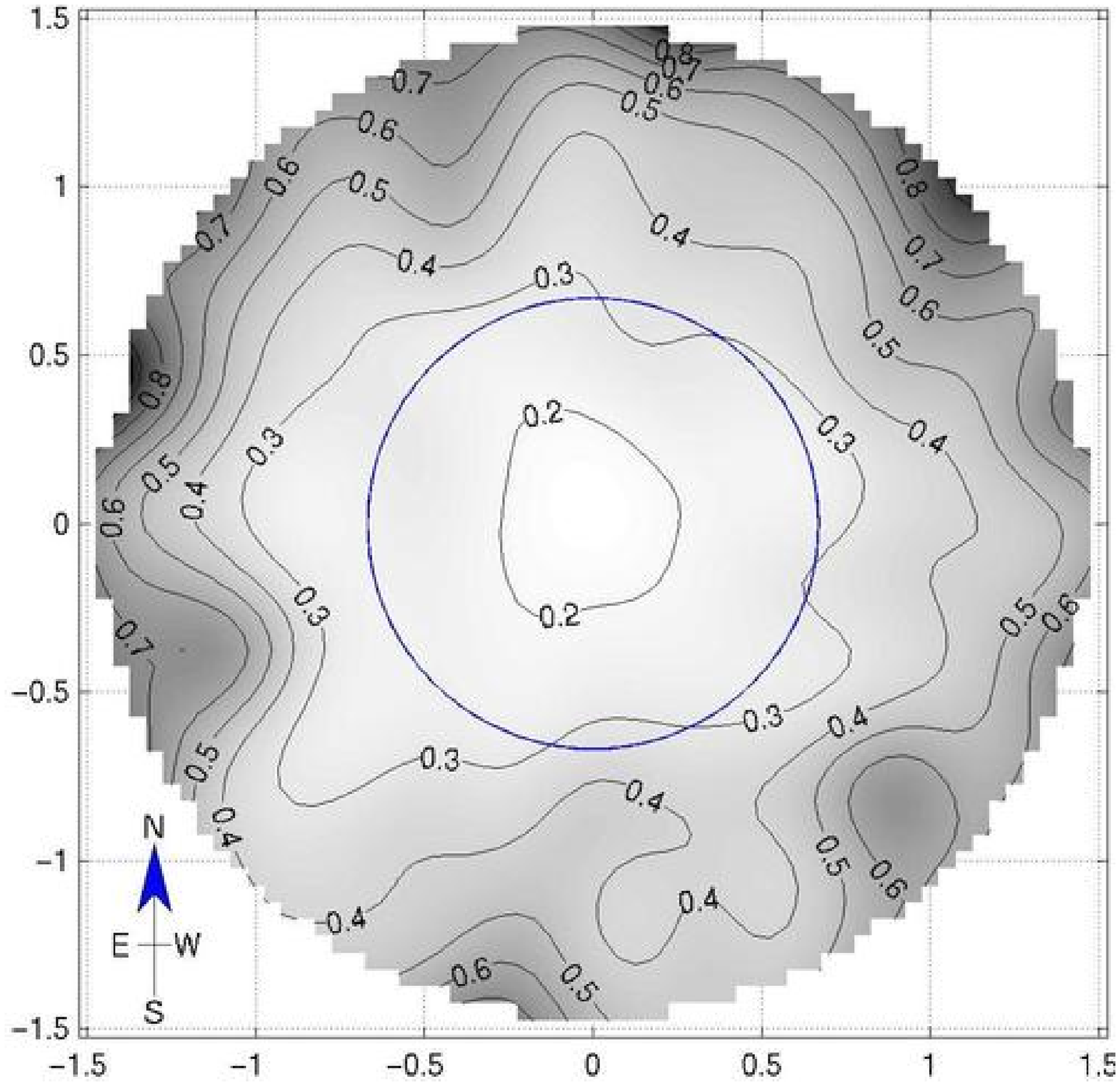}}
\end{minipage}
\begin{minipage}{\plotwidth}
\centerline{3EG J1323+2200}
\resizebox{\textwidth}{!}{\includegraphics{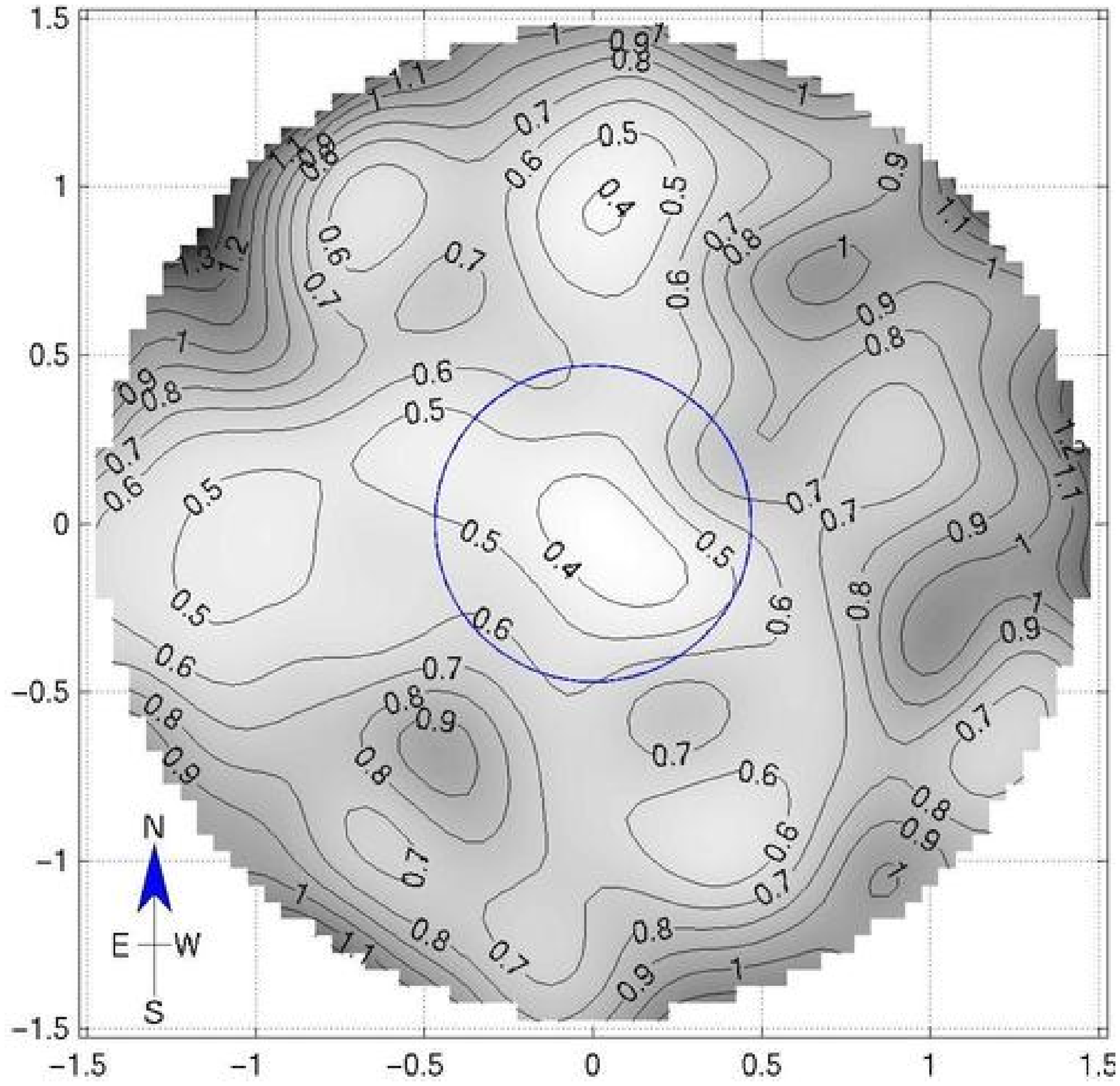}}
\end{minipage}
\begin{minipage}{\plotwidth}
\centerline{3EG J1907+0556}
\resizebox{\textwidth}{!}{\includegraphics{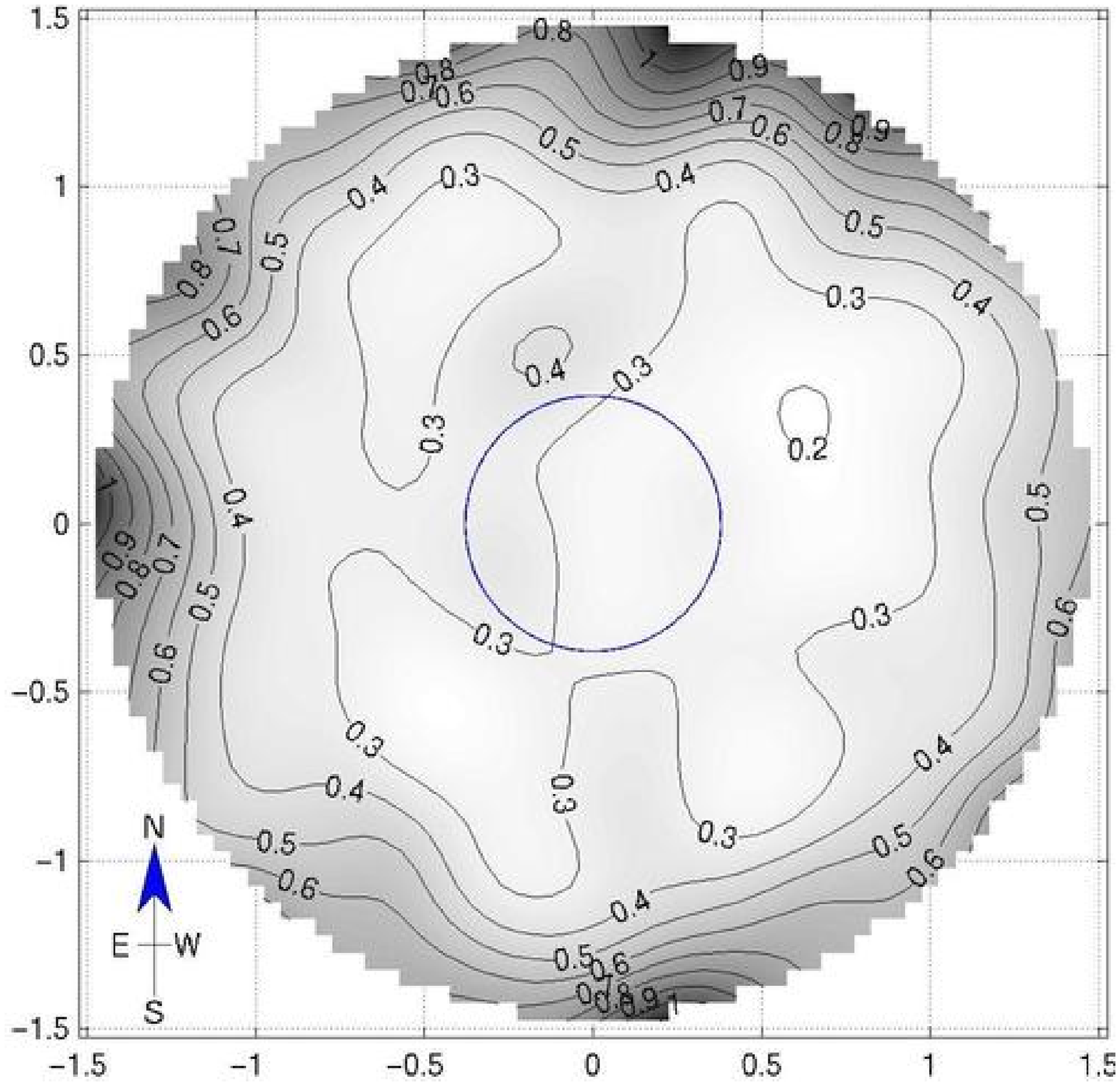}}
\end{minipage}

\vspace*{2ex}
\begin{minipage}{\plotwidth}
\centerline{3EG J2020+3658}
\resizebox{\textwidth}{!}{\includegraphics{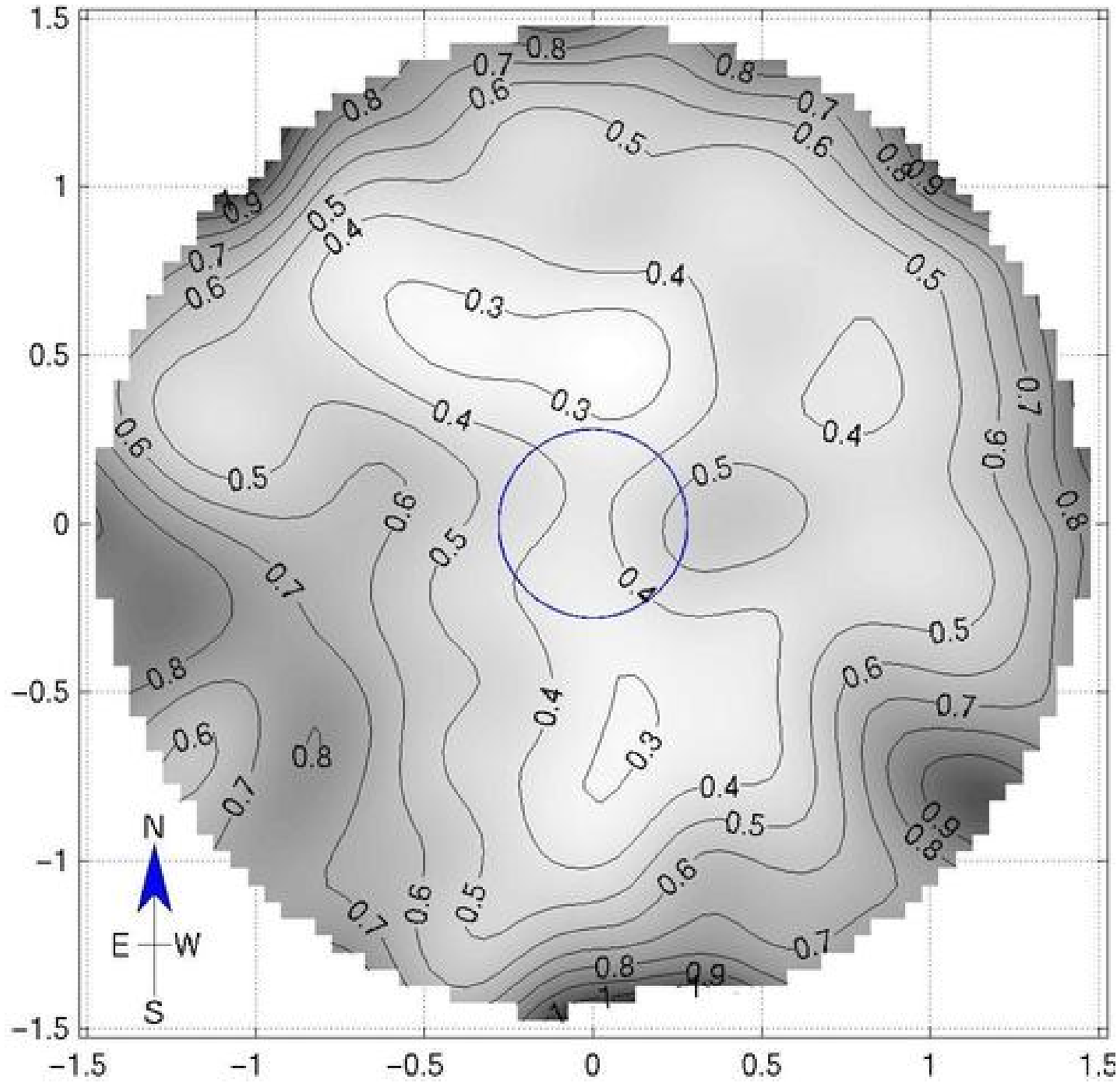}}
\end{minipage}
\begin{minipage}{\plotwidth}
\centerline{3EG J2227+6122}
\resizebox{\textwidth}{!}{\includegraphics{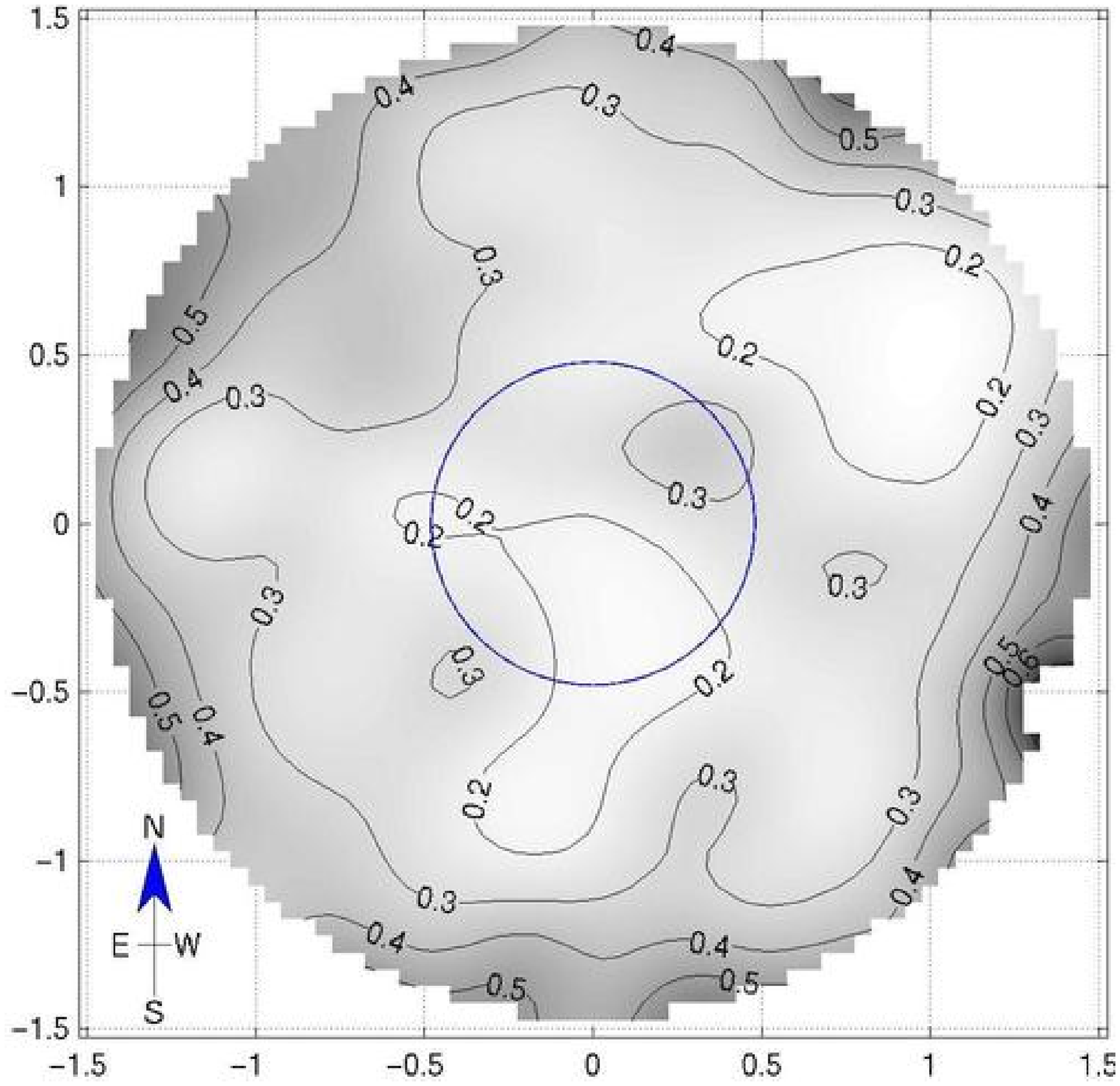}}
\end{minipage}
\end{minipage}

\caption{\label{RESULTS::UPPERLIMITS} VHE upper limits on emission. 
Upper limits are given in units of the Crab Nebula flux. The axes are
in degrees from the center of the field of view as given in Table
\ref{OBS::SOURCES}. Increasing declination is toward the top of each
plot, increasing RA to the left. The circle indicates the 95\%
confidence circle from \citet{Lamb97} or \citet{Hartman99} as
appropriate. Where error ellipses have been given in in
\citet{Lamb97}, a circle with radius equal to the semi-major axis is
displayed. The maximum upper limit in each 95\% confidence region is
given in Table \ref{RESULTS::TABLE}.}
\end{figure}

In one case, GeV J1907+0556, the analysis indicated significant
emission throughout the 7 square degree field, the result of large
brightness differences between ON and OFF observations that was not
fully compensated for in padding. For this source alone, the ON source
counts were scaled by a value calculated by examining the number of
counts in the region of $1.4^\circ<r<1.8^\circ$ from the center of the
field of view.

No significant emission was detected from any source. Figure
\ref{RESULTS::UPPERLIMITS} shows upper limits on emission from the
sources observed. Table \ref{RESULTS::TABLE} summarizes these results
for the error circle of each object. In each case, the highest limit
found in each region is quoted.

\begin{table}[t]
\caption{\label{RESULTS::TABLE} Upper limits derived from Figure
\ref{RESULTS::UPPERLIMITS}.}
\begin{tabular}{ccc}
 & \emphasise{Positional Error\tablenote{95\% confidence circle from \citet{Lamb97} or \citet{Hartman99} as appropriate.}} & \emphasise{Upper Limit \tablenote{Fluxes in units of $10^{-11}\mathrm{cm}^{-2}\mathrm{s}^{-1}$ calculated
from measured Crab flux of \citet{Hillas98}.}} \\ 
\emphasise{Source} & \emphasise{(degrees)} & 
\emphasise{(\boldmath $\mathbf{E}>$430GeV)} \\ \hline
3EG J0423+1707 & 0.77 & 3.6 \\ 
GeV J0433+2907 & 0.35 & 2.1 \\
3EG J0450+1105 & 0.64 & 3.8 \\
3EG J0634+0521 & 0.67 & 2.4 \\
3EG J1323+2200 & 0.47 & 5.9 \\
GeV J1907+0556 & 0.38 & 2.7 \\
GeV J2020+3658 & 0.28 & 4.1 \\
3EG J2227+6122 & 0.48 & 2.6 \\
\end{tabular}
\end{table}

%

%


\end{document}